\documentclass[11pt]{article}
\usepackage[T1]{fontenc}
\usepackage[utf8]{inputenc}
\usepackage{float}
\usepackage{amsmath}
\usepackage{amssymb}
\usepackage{graphicx}
\usepackage[authoryear]{natbib}
\usepackage[unicode=true,
 bookmarks=false,
 breaklinks=false,pdfborder={0 0 1},backref=section,colorlinks=false]
 {hyperref}

\makeatletter

\providecommand{\tabularnewline}{\\}
\floatstyle{ruled}
\newfloat{algorithm}{tbp}{loa}
\providecommand{\algorithmname}{Algorithm}
\floatname{algorithm}{\protect\algorithmname}

\usepackage{algolyx}

\@ifundefined{date}{}{\date{}}
\usepackage{latexsym}
\usepackage{multirow}\usepackage{bm}%\usepackage[latin9]{inputenc}
\usepackage{url}% not crucial - just used below for the URL 

\usepackage{enumerate}

\usepackage{colortbl}
\usepackage{subcaption}

\usepackage{dcolumn}
\newcolumntype{.}{D{.}{.}{-1}}
\newcolumntype{d}[1]{D{.}{.}{#1}}
\usepackage{theorem}
\newtheorem{assumption}{Assumption}\newtheorem{lemma}{Lemma}

\newcommand{\ind}{\mbox{$\perp\!\!\!\perp$}}

\usepackage{rotating}

\usepackage{arydshln}

\usepackage[compact]{titlesec}

\allowdisplaybreaks

\newcommand{\spacingset}[1]{\renewcommand{\baselinestretch}%
{#1}\small\normalsize}

\newcommand{\T}{\mathrm{\scriptscriptstyle T}}

\makeatother

\begin{document}
\title{\textbf{Estimating optimal interpretable individualized treatment
regimes from a classification perspective using adaptive LASSO}}
\author{Yunshu Zhang, Shu Yang, Wendy Ye, Ilya Lipkovich, Douglas Faries}

\maketitle

\spacingset{1.5} 
\begin{abstract}
Real-world data (RWD) gains growing interests to provide a representative
sample of the population for selecting the optimal treatment options. However,
existing complex black box methods for estimating individualized
treatment rules (ITR) from RWD have problems in interpretability and
convergence. Providing an interpretable and sparse ITR can be used
to overcome the limitation of existing methods. We developed an algorithm
using Adaptive LASSO to predict optimal interpretable linear ITR in
the RWD. To encourage sparsity, we obtain an ITR by minimizing the
risk function with various types of penalties and different methods
of contrast estimation. Simulation studies were conducted to select
the best configuration and to compare the novel algorithm with
the existing state-of-the-art methods. The proposed algorithm was
applied to RWD to predict the optimal interpretable ITR. Simulations
show that adaptive LASSO had the highest rates of correctly selected
variables and augmented inverse probability weighting with Super Learner
performed best for estimating treatment contrast. Our method had a 
better performance than causal forest and R-learning
in terms of the value function and variable selection. The proposed algorithm
can strike a balance between the interpretability of estimated ITR (by selecting a small set of
important variables) and its value.
\end{abstract}
\newpage{}

\section{Introduction}

Within the causal inference literature, researchers have dedicated significant attention to estimating the average  treatment effects the average treatment effect in the overall population (ATE), and the average treatment effect in the treated (ATT). Nevertheless, given the potential existence of heterogeneity in treatment effects across both clinical trials and observational studies, it becomes imperative to transcend the confines of ATE and ATT. There arises a compelling need to explore beyond these averages. The prospect of designing individualized treatment regimes (ITR) or pinpointing subgroups that exhibit a higher efficacy in response to the treatment (compared to teh overall population) becomes a pertinent avenue to explore. \citet{lipkovich2017tutorial} present a comprehensive review encompassing this overarching framework.

In the realm of statistical methods employed for subgroup discovery, the pivotal step often involves estimating individualized treatment effects (ITE), denoted as $\tau\left(x\right)$, or equivalently, the contrast function. Various approaches come into play for this estimation. Methods such as univariate regression or tree-based regression models (e.g., CART \citep{breiman2017classification}) are employed to estimate outcome functions for both treatment arms, incorporating treatment-by-biomarker interactions where applicable. When dealing with a substantial number of covariates, the use of penalized regression techniques (e.g., LASSO \citep{tibshirani1996regression} or the elastic net \citep{zou2006adaptive}), or black box models (e.g., random forest \citep{breiman2001random}), becomes necessary to tackle the complexity of estimation. Alternatively, there exists a methodology wherein the estimation of ITE occurs directly, without estimating the main effects. This approach involves global direct treatment effect modeling methods, such as GUIDE \citep{loh2015regression}, causal Bayesian trees \citep{hahn2020bayesian}, and R-learning \citep{nie2021quasi}. Once the ITE is derived, the selection of subgroups often involves criteria such as $\left\{ x:\hat{\tau}\left(x\right)>0\right\}$ or $\left\{ x:\hat{\tau}\left(x\right)>\delta\right\}$.

\begin{figure}
\caption{Illustrating example comparing direct method and indirect method.
The true ITE is a second-degree polynomial function of $x$, but the true
optimal ITR is linear. The indirect method is concerned with minimizing the
prediction error, while the direct method focuses on the sign and thus
approaches the true optimal regime by-passing contrast estimation. \label{fig:Illustrating-example}}

\begin{centering}
\includegraphics[width=\textwidth]{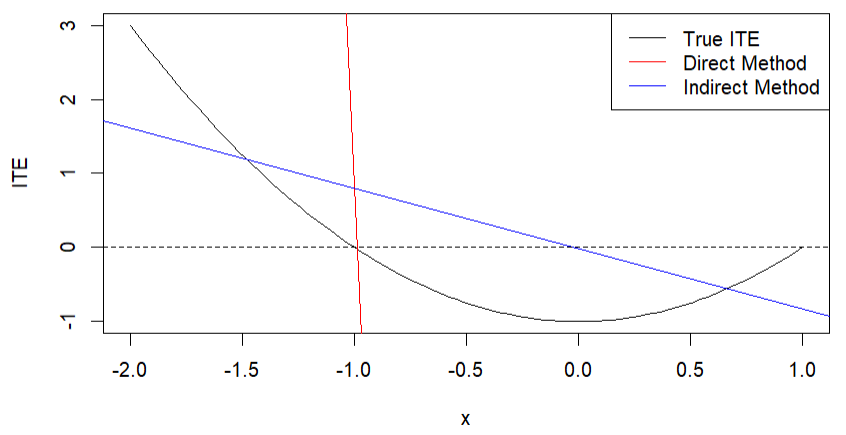}
\par\end{centering}
\end{figure}

While obtaining the Individualized Treatment Effect (ITE) $\tau\left(x\right)$ is adequate for deriving the optimal Individualized Treatment Regime (ITR), it's not a necessary precursor. Only the sign of the contrast function holds significance, rendering the complete estimation of $\tau\left(x\right)$ unnecessary. This task is challenging due to the potentially complex nature of the contrast function, often requiring algorithms that use reduced models, such as linear models. However, the ITE might not follow a linear pattern even if the optimal ITR does, as illustrated in Figure \ref{fig:Illustrating-example}. Hence, a more effective approach might involve directly modeling the ITR rather than indirectly estimating the ITE.

One avenue involves maximizing the value function (defined in Section~\ref{sec:Notation-and-assumptions}), which gauges the expected outcome when subjects receive treatments following a specified treatment regime. However, estimating the value function hinges on missing potential outcomes (counterfactual outcomes), necessitating estimation from observed data. Various methods have been proposed for this purpose, including inverse probability weighting (IPW) \citep{horvitz1952generalization}, outcome regression (OR) \citep{murphy2003optimal}, and augmented inverse probability weighting (AIPW) \citep{zhang2012robust}.

The challenge further lies in optimizing the value function, a non-convex and non-standard function that is intricate to maximize. While grid search or genetic algorithms have been used for this purpose \citep{zhang2012estimating}, they often fail in high-dimensional settings. An alternate method frames this as a classification problem, as highlighted by \citet{bai2017optimal}: finding the optimal ITR is akin to minimizing a risk function. Algorithms like outcome weighted learning \citep{zhao2012estimating} and CAPITAL \citep{cai2022capital} embrace this concept. Nevertheless, the optimization remains challenging due to the non-convex nature of the risk function. Researchers have explored smooth surrogate functions to overcome this obstacle \citep{zhou2017residual,bai2017optimal,wu2023transfer}.

In real-world applications, ensuring the interpretability of the chosen subgroup is crucial. Interpretability here encompasses two key aspects: the structure of the treatment regime and the number of variables involved. While complex models like random forests and neural networks often yield higher efficiency, advocating treatments without easily understandable explanations is untenable. Consequently, researchers lean towards employing linear rules or decision trees to construct treatment regimes that are more interpretable.

Furthermore, reducing the number of covariates included in the policy contributes significantly to enhancing interpretability. Simple regularization techniques like LASSO may be integrated to streamline the policy \citep{bai2017optimal}. However, these methods might lack oracle properties, such as selection consistency \citep{zou2006adaptive}. Advanced variable selection techniques, like adaptive LASSO, remain underexplored concerning linear Individualized Treatment Regimes (ITR).

In tree-based methodologies, methods like Virtual Twins (VT) \citep{foster2011subgroup} aid in simplifying the algorithm by pruning the tree and retaining only the crucial covariates. However, as a multi-stage procedure it may be suboptimal. Alternatively, defining variable importance and directly selecting prognostically crucial variables has been proposed \citep{williamson2021discussion}. These variables are significantly linked to potential outcomes. Yet, they might differ from the set of covariates important solely for predicting Individualized Treatment Effects (ITE) \citep{lipkovich2017tutorial}. 

This paper introduces a novel algorithm designed to estimate an optimal and interpretable Individualized Treatment Regime (ITR) in high-dimensional settings. Here, ``optimal'' refers to maximizing the value function, while ``interpretable'' pertains to a linear policy restricted to a limited number of predictors. Adopting the classification perspective highlighted by \citet{bai2017optimal}, we integrate adaptive LASSO into our algorithm for effective variable selection. Given the inapplicability of grid search and genetic algorithms in high-dimensional scenarios, we utilize two surrogate functions\textemdash the smoothed ramp loss function and the convex hinge loss function\textemdash to ensure computational feasibility. These functions correspond to the weighted support vector machine and the d.c. algorithm \citep{thi1997solving}, respectively. Our methodology employs cross-validation to select tuning parameters, followed by a recommended refitting process. Additionally, we provide a complementary analysis procedure for flexible variable selection, aiding researchers in choosing more practical and efficient policies. Furthermore, we extend our algorithm to accommodate survival outcomes (shown in the appendix). To illustrate our method, we apply the algorithm to the TRIUMPH dataset for migraine \citep{lipton2025treatment}. Simulation studies are conducted to compare the performance of our algorithm against existing state-of-the-art methods.

The remaining part of this paper is organized as follows: Section
\ref{sec:Notation-and-assumptions} introduces the basic notation
and assumptions. Section \ref{sec:algorithm} describes the details
of our proposed algorithm. Section \ref{sec:Realdata}
applies the algorithm to the real-world dataset. Section \ref{sec:simulation} uses simulation studies to illustrate our algorithm.  Section \ref{sec:Discussion} concludes the article and discusses potential limitations and future research.

\section{Notation and assumptions\label{sec:Notation-and-assumptions} }

Let $X_{i}\in\mathcal{X}$ be a $p$-dimensional vector of pre-treatment
covariates, $A_{i}\in\left(0,1\right)$ be the binary treatment, and
$Y_{i}\in\mathbb{R}$ be the outcome for unit $i=1,\ldots,n$. We
denote $A_{i}=1$ as the positive treatment or treatment group and
$A_{i}=0$ as the negative treatment or control group. The outcome
is allowed to be binary, and we assume that a higher outcome is desired,
for example, the percentage of improvement of a patient's health measurement.
We follow the potential outcomes framework. Let $Y_{i}\left(a\right)$
be the potential outcome had unit $i$ been given treatment $a$ $\text{\ensuremath{\left(a=0,1\right)}}$.
Based on the potential outcomes, the ATE is $\tau=\mathbb{E}\left\{ Y_{i}\left(1\right)-Y_{i}\left(0\right)\right\} $
and the ATT is $\tau_{\mathrm{ATT}}=\mathbb{E}\left\{ Y_{i}\left(1\right)-Y_{i}\left(0\right)\mid A_{i}=1\right\} $.
In this paper, we are interested in estimating the ITE $\tau\left(x\right)=\mathbb{E}\left\{ Y_{i}\left(1\right)-Y_{i}\left(0\right)\mid X_{i}=x\right\} $,
or equivalently, the contrast function. The observed outcome is $Y_{i}=Y_{i}\left(A_{i}\right)=A_{i}Y_{i}\left(1\right)+\left(1-A_{i}\right)Y_{i}\left(0\right)$.
We assume that $\left\{ X_{i},A_{i},Y_{i}\left(0\right),Y_{i}\left(1\right)\right\} $,
$i=1,\ldots,n$, are independent and identically distributed. Thus,
$\left(X_{i},A_{i},Y_{i}\right)$, $i=1,\ldots,n$, are also independent
and identically distributed. 

To identify the causal effects, we make the standard ``no unmeasured confounders''
and the positivity assumptions \citep{rosenbaum1983central}.

\begin{assumption}\label{asump-nuc} (No unmeasured confounder) The potential outcomes are conditionally
independent with the treatment assignment given the observed covariates:
$Y\left(a\right)\ind A\mid X$.

\end{assumption}

\begin{assumption}\label{asump-overlap} (Positivity) There exist constants $c_{1}$
and $c_{2}$ such that $0<c_{1}\leq e\left(X\right)\leq c_{2}<1$
almost surely, where $e\left(X\right)=\mathbb{P}\left(A=1\mid X\right)$
is the propensity score as the probability to receive positive treatment. 

\end{assumption}

A treatment regime or a policy $d\left(x\right)$ is defined as a
function from the covariate space $\mathcal{X}$ to the treatment
indicators $\left(0,1\right)$. If $d\left(x\right)=1$, the patient
with baseline covariates $X=x$ would receive the treatment $1$.
Similarly, the treatment 0 would be assigned to the patient if $d\left(x\right)=0$.
To evaluate a policy, we define the value function as the expected
outcome if the treatment assignments are assigned following the treatment
regime $d$: $V\left(d\right)=\mathbb{E}\left[Y\left( d\left(X\right)\right) \right]$.
Because we assume a higher outcome is beneficial, the optimal policy $d^{opt}$ is
defined to maximize the value function: $V\left(d^{opt}\right)=\textrm{max}_{d\in\mathcal{D}}V\left(d\right)$,
where $\mathcal{D}$ is the space of all possible treatment regimes. 

Because value function depends on the missing potential outcomes,
it is necessary to estimate the value function based on the observed
data. Existing methods include inverse probability weighting (IPW)\citep{horvitz1952generalization}
and outcome regression (OR)\citep{murphy2003optimal}, where the first
method relies on the propensity score $e\left(X\right)$ and the second
method depends on the expected potential outcome or the Q-function
$\mu\left(X;a\right)=\mathbb{E}\left\{ Y\left(a\right)\mid X\right\} $.
Both parametric and non-parametric algorithms have been used to estimate
these nuisance functions. However, IPW and OR estimators are inconsistent
when the corresponding model is not correctly specified. To improve
the robustness of the estimate, \citet{zhang2012robust} proposed
the augmented inverse probability weighting (AIPW) estimator:
\begin{eqnarray}
\hat{V}_{aipw}\left(d\right) & = & \frac{1}{n}\sum_{i=1}^{n}\left(\left[\frac{A_{i}d\left(X_{i}\right)}{\hat{e}\left(X_{i}\right)}+\frac{\left(1-A_{i}\right)\left\{ 1-d\left(X_{i}\right)\right\} }{1-\hat{e}\left(X_{i}\right)}\right]\left[Y_{i}-\hat{\mu}\left\{ X_{i};d\left(X_{i}\right)\right\} \right]+\left[Y_{i}-\hat{\mu}\left\{ X_{i};d\left(X_{i}\right)\right\} \right]\right).\label{eq:valueAIPW}
\end{eqnarray}
The AIPW estimator is consistent to the true value of the treatment
regime if either the propensity score or outcome model is correctly
speicified, which is the so-called doubly robust property. Another
important property of the AIPW estimator is its semiparametric efficiency
in the sense that its asymptotical variance is the smallest in the
class of semiparametric estimators for the value function, and the
asymptotical variance can be estimated via the influence function
following \eqref{eq:valueAIPWvar}. Optimization algorithms can then
be applied to the estimated value function to search for the optimal
ITR.
\begin{equation}
\begin{aligned}
\mathbb{V}\!\big(\hat{V}_{\mathrm{aipw}}(d)\big)
&= \frac{1}{n^2}\sum_{i=1}^{n}\Big(
\Big[\tfrac{A_i d(X_i)}{\hat e(X_i)}+\tfrac{(1-A_i)\{1-d(X_i)\}}{1-\hat e(X_i)}\Big]\big[Y_i-\hat{\mu}\{X_i; d(X_i)\}\big] \\
&\qquad\qquad\qquad\quad + Y_i - \hat{\mu}\{X_i; d(X_i)\} - \hat{V}_{\mathrm{aipw}}(d)
\Big)^2 .
\end{aligned}
\label{eq:valueAIPWvar}
\end{equation}

While sophisticated nonlinear algorithms can derive policies with greater value, their complexity often sacrifices interpretability. Particularly, for opaque methods like neural networks, patients and physicians might hesitate to accept treatment recommendations lacking understandable reasoning. Consequently, researchers often prioritize interpretability, accepting a trade-off in the value function. Linear and tree-based rules have garnered significant attention in the literature. This paper focuses on a specific class of constrained linear regimes, denoted as $\mathcal{D}_{\eta}$, wherein each policy adheres to a linear rule: $d_{\eta}\left(x\right)=d\left(x;\eta\right)=I\left(x^{\T}\eta>0\right)$. Here, $\eta\in\mathbb{R}^{p}$ represents the coefficient for this linear policy. To streamline the discussion, we assume the intercept term is already incorporated in $X$, obviating the need for a threshold term $c$ in defining the linear policy. An optimal linear regime $d_{\eta}^{opt}$, characterized by the coefficient $\eta^{opt}$, maximizes the value function among all linear regimes $d_{\eta}\in\mathcal{D}_{\eta}$, where $\eta^{opt}=\textrm{argmax}{\eta}V\left(d_{\eta}\right)$. The primary objective of this paper is to estimate this optimal interpretable individualized treatment regime $d_{\eta}^{opt}$.

Different from the standard generalized linear models such as logistic
and probit regression, the linear score $f\left(x;\eta\right)=x^{\T}\eta$
cannot be directly interpreted to be related to the probability of
a patient with baseline covariate $x$ to receive positive treatment.
The rigorous interpretation is that: for a positive coefficient, if
we compare patients with larger number of the corresponding covariate
versus patients with smaller number of the covariate, the first population
has a greater proportion to receive the treatment. The magnitude of
the coefficient determines the difference between the proportions.
A easier interpretation is that: a large positive coefficient makes
the patient with larger corresponding covariate more ``likely''
to get a recommendation of positive treatment. 

\section{Algorithm \label{sec:algorithm}}

\subsection{Classification perspective and surrogate functions\label{subsec:surrogate}}

By definition, it is adequate to find the optimal linear regime $d_{\eta}^{opt}$
by maximizing the value function $V\left(d_{\eta}\right)$ in the
restricted class of linear regimes $\mathcal{D}_{\eta}$. However,
it is a non-standard optimization problem because $V\left(d_{\eta}\right)$
is a non-convex and non-smooth function of $\eta$. Classical convex
optimization algorithms cannot be applied to the problem. Researchers
have been using non-convex value search algorithms to find the maximizer,
for example, the grid search or genetic algorithm \citep{zhang2012estimating}.
But these types of algorithms are not applicable in high dimensional
settings due to the unaffordable computational burden. To deal with
this problem, \citet{bai2017optimal} reconsider this problem from
a classification perspective, as shown in the following lemma.

\begin{lemma}\label{lemma-risk} \citep{zhang2012estimating} The
coefficient of the optimal ITR $\eta^{opt}$, which maximizes the
value function $V\left(d_{\eta}\right)$, is also the minimizer of
the risk function 
\begin{eqnarray*}
\mathcal{R}_{\mathcal{F}}\left(\eta;\tau,l_{0-1}\right) & = & \mathbb{E}\left\{ \left|\tau\left(X\right)\right|l_{0-1}\left(\left[2I\left\{ \tau\left(X\right)>0\right\} -1\right]f\left(X;\eta\right)\right)\right\} \\
 & = & \mathbb{E}\left(\left|\tau\left(X\right)\right|\left[I\left\{ \tau\left(X\right) > 0 \right\} -d\left(X;\eta\right) \right]^{2}\right),
\end{eqnarray*}
where $l_{0-1}\left(u\right)=I\left(u\leq0\right)$ and $\mathcal{F}$
is the domain of the ITR coefficient $\eta$, for example, $\mathbb{R}^{p}$. 

\end{lemma}

The lemma illustrates the definition of optimal ITR from a weighted
classification perspective. If we know the ITE $\tau\left(x\right)$,
by definition, the optimal ITR assigns subjects based on the sign of
$\tau\left(x\right)$: $d^{opt}\left(x\right)=I\left(\tau\left(x\right)>0\right)$.
Thus, it is reasonable to evaluate a policy by comparing its treatment
assignments with the optimal ITR. If there is a disagreement, $\left[I\left\{ \tau\left(X\right) > 0 \right\} -d\left(X;\eta\right) \right]^{2}=1$
and it adds up to the risk function, where the magnitude is based
on the weight $\left|\tau\left(X\right)\right|$. For patients with
larger differences in their potential outcomes, it is more risky to
make mistakes in their treatment assignments. In practice, because
of the missing potential outcomes, the contrast function $\tau\left(X\right)$
needs to be estimated from the observed data. For example, we can
use the AIPW type estimator to estimate the ITE:
\begin{eqnarray}
\hat{\tau}_{aipw}\left(X_{i}\right) & = & \frac{A_{i}\left\{ Y_{i}-\hat{\mu}\left(X_{i};1\right)\right\} }{\hat{e}\left(X_{i}\right)}-\frac{\left(1-A_{i}\right)\left\{ Y_{i}-\hat{\mu}\left(X_{i};0\right)\right\} }{1-\hat{e}\left(X_{i}\right)}+\hat{\mu}\left(X_{i};1\right)-\hat{\mu}\left(X_{i};0\right).\label{eq:iteAIPW}
\end{eqnarray}
Similar to $\hat{V}_{aipw}\left(d\right)$, the AIPW estimator of
the ITE also enjoys the doubly robust property. In practice, we replace
$\tau\left(X_{i}\right)$ by $\hat{\tau}\left(X_{i}\right)$ and minimize
the following empirical risk function to find $\hat{\eta}^{opt}$
\begin{eqnarray*}
\hat{\mathcal{R}}_{\mathcal{F}}\left(\eta;\hat{\tau},l_{0-1}\right) & = & \frac{1}{n}\sum_{i=1}^{n}\left\{ \left|\hat{\tau}\left(X_{i}\right)\right|l_{0-1}\left(\left[2I\left\{ \hat{\tau}\left(X_{i}\right)>0\right\} -1\right]f\left(X_{i};\eta\right)\right)\right\} \\
 & = & \frac{1}{n}\sum_{i=1}^{n}\left(\left|\hat{\tau}\left(X_{i}\right)\right|\left[I\left\{ \hat{\tau}\left(X_{i}\right)-d\left(X_{i};\eta\right)\right\} \right]^{2}\right).
\end{eqnarray*}

However, even the risk function is a non-convex and non-standard function
of $\eta$. The difficulty comes from the nature of the 0-1 loss function
$l_{0-1}\left(u\right)=I\left(u\leq0\right)$, which is non-continuous
and non-differentiable at $0$ and non-convex in its domain. To deal
with this problem, researchers have proposed to replace the 0-1 loss
function with other loss functions, which leads to different algorithms.
The first choice is the convex Hinge loss function $l_{h}\left(u\right)=\textrm{max}\left(1-u,0\right)$
\citep{bai2017optimal}, and it turns out to be a weighted support
vector machine (WSVM) formulation \citep{yang2005weighted}. This
is a convex optimization problem that can be solved computationally
easily. Note that the outcome weighted learning \citep{zhao2012estimating}
estimator also applies this convex Hinge loss function but uses the
IPW-based ITE estimates instead of $\hat{\tau}_{aipw}$. To obtain
better robustness and efficiency, we focus on the WSVM in this paper.
The objective function to be minimized is
\begin{eqnarray*}
\hat{\mathcal{R}}_{\mathcal{F}}\left(\eta;\hat{\tau},l_{h}\right) & = & \frac{1}{n}\sum_{i=1}^{n}\left\{ \left|\hat{\tau}\left(X_{i}\right)\right|l_{h}\left(\left[2I\left\{ \hat{\tau}\left(X_{i}\right)>0\right\} -1\right]f\left(X_{i};\eta\right)\right)\right\} \\
 & = & \frac{1}{n}\sum_{i=1}^{n}\left(\left|\hat{\tau}\left(X_{i}\right)\right|\textrm{max}\left(1-\left[2I\left\{ \hat{\tau}\left(X_{i}\right)>0\right\} -1\right]f\left(X_{i};\eta\right),0\right)\right).
\end{eqnarray*}

Another choice is the smoothed ramp loss function $l_{r}\left(u\right)$
\citep{zhou2017residual,wu2023transfer} defined as follows, which
benefits from being smooth everywhere and robust to outliers. 
\begin{center}
$l_{r}\left(u\right)=\begin{cases}
0 & \textrm{if }u\geq1,\\
\left(1-u\right)^{2} & \textrm{if }0\leq u<1,\\
2-\left(1+u\right)^{2} & \textrm{if }-1\leq u<0,\\
2 & \textrm{if }u\leq-1.
\end{cases}$ $l_{s}\left(u\right)=\begin{cases}
0 & \textrm{if }u\geq s,\\
\left(s-u\right)^{2} & \textrm{if }s-1\leq u<s,\\
2s-2u-1 & \textrm{if }u<s-1.
\end{cases}$
\par\end{center}

Then, the risk function to be minimized is 
\begin{eqnarray*}
\hat{\mathcal{R}}_{\mathcal{F}}\left(\eta;\hat{\tau},l_{r}\right) & = & \frac{1}{n}\sum_{i=1}^{n}\left\{ \left|\hat{\tau}\left(X_{i}\right)\right|l_{r}\left(\left[2I\left\{ \hat{\tau}\left(X_{i}\right)>0\right\} -1\right]f\left(X_{i};\eta\right)\right)\right\} \\
 & = & \frac{1}{n}\sum_{i=1}^{n}\left\{ \left|\hat{\tau}\left(X_{i}\right)\right|l_{1}\left(\left[2I\left\{ \hat{\tau}\left(X_{i}\right)>0\right\} -1\right]f\left(X_{i};\eta\right)\right)\right\} \\
 &  & -\frac{1}{n}\sum_{i=1}^{n}\left\{ \left|\hat{\tau}\left(X_{i}\right)\right|l_{0}\left(\left[2I\left\{ \hat{\tau}\left(X_{i}\right)>0\right\} -1\right]f\left(X_{i};\eta\right)\right)\right\} .
\end{eqnarray*}
The function also has an important property that $l_{r}\left(u\right)$
can be decomposed into the difference of two convex functions $l_{1}\left(u\right)$
and $l_{0}\left(u\right)$, where $l_{s}\left(u\right)$ is defined
as a piecewise polynomial function as well. It is important because
we can apply the d.c. algorithm \citep{thi1997solving} to solve this
non-convex optimization problem iteratively as pointed out by \citet{zhou2017residual}.
The detail of the algorithm will be introduced in Section \ref{subsec:Smoothed-ramp-loss}
after we introduce the penalizing term in the following section. 

\subsection{Penalize the coefficients via LASSO and adaptive LASSO\label{subsec:Penalize}}

To enhance out-of-sample predictive performance and mitigate overfitting, researchers have introduced various penalizing terms into the objective function. A commonly employed strategy involves incorporating the norm of the policy coefficient multiplied by a regularization parameter. Notable instances include LASSO, utilizing the L1 norm \citep{tibshirani1996regression}, and Ridge regression, utilizing the L2 norm \citep{hoerl1970ridge}. In comparing these methods, LASSO stands out for its capacity to reduce dimensions and select crucial variables, thereby enhancing the interpretability of the policy.

The penalty term in LASSO is defined as:
\[
J_{lasso}\left(\eta\right)=\lambda\sum_{j=1}^{p}\left|\eta_{j}\right|.
\]
Here, $\lambda$ represents the regularization parameter, typically selected through cross-validation. However, it's worth noting that LASSO fails to attain the oracle property \citep{fan2001variable}, which encompasses two crucial aspects: selection consistency and estimation consistency. Achieving this property led to the proposal of adaptive LASSO by \citet{zou2006adaptive}. In adaptive LASSO, regularization parameters vary for different coefficients, introducing an initial estimate $\hat{\eta}_{\text{int}}$. The penalty term in adaptive LASSO is given by:
\[
J_{adplasso}\left(\eta;\hat{\eta}_{int}\right)=\lambda\sum_{j=1}^{p}\left(\left|\eta_{j}\right|/\left|\hat{\eta}_{int,j}^{\gamma}\right|\right).
\]
Here, $\gamma$ is a positive tuning parameter regulating the impact from the initial estimate. Although \citet{zou2006adaptive} suggested a two-dimensional cross-validation for tuning adaptive LASSO, in this paper, for computational efficiency, we opt to fix $\gamma$ as a constant. For example, setting $\gamma = 1$ aligns closely with the nonnegative garotte algorithm proposed by \citet{breiman1995better}. The initial estimate $\hat{\eta}_{\text{int}}$ can be obtained from the LASSO estimator with $\lambda=0$, which is the standard ordinary least square estimator with a convergence rate of $\sqrt{n}$, satisfying the requirements of adaptive LASSO \citep{zou2006adaptive}.

It's important to note that the risk function $\mathcal{R}{\mathcal{F}}(\eta;\tau,l{0-1})$ remains invariant to the scale of $\eta$, meaning the objective function remains unchanged when all coefficients scale down to zero at the same rate. Consequently, $\mathcal{R}{\mathcal{F}}(\eta;\tau,l{0-1})$ possesses an infinite number of minimizers, with the regularization term favoring the smallest among them. To prevent $\hat{\eta}$ from becoming exceedingly small, alternatives to the penalty terms can be considered, such as fixing one non-zero coefficient as a constant. However, while $\mathcal{R}{\mathcal{F}}(\eta;\tau,l{0-1})$ remains unchanged with uniform changes in $\eta$, the surrogate functions $\mathcal{R}{\mathcal{F}}(\eta;\tau,l{h})$ and $\mathcal{R}{\mathcal{F}}(\eta;\tau,l{r})$ do not maintain this scale invariance. Consequently, these practical risk functions possess unique minimizers, ensuring that regularization terms don't yield extreme solutions. In simulations outlined in the appendix, we discovered that the standard adaptive LASSO algorithm performs equally well compared to other variants. Therefore, we recommend directly employing the adaptive LASSO penalty $J_{\text{adplasso}}$ within the ITR algorithm.

\subsection{Smoothed ramp loss function and d.c. algorithm\label{subsec:Smoothed-ramp-loss}}

Combining the surrogate loss function in Setion \ref{subsec:surrogate}
and the penalty term in Section \ref{subsec:Penalize}, we minimize
the penalized loss function to obtain the optimal interpretable ITR.
For example, if hinge loss function is used to replace the 0-1 loss
function, $\hat{\eta}_{wsvm}=\textrm{argmin}_{\eta}\hat{\mathcal{L}}_{\mathcal{F}}\left(\eta;\hat{\tau},\hat{\eta}_{int},l_{h}\right)$,
where
\begin{eqnarray}
\hat{\mathcal{L}}_{\mathcal{F}}\left(\eta;\hat{\tau},\hat{\eta}_{int},l_{h}\right) & = & \hat{\mathcal{R}}_{\mathcal{F}}\left(\eta;\hat{\tau},l_{h}\right)+J_{adplasso}\left(\eta;\hat{\eta}_{int}\right)\nonumber \\
 & = & \frac{1}{n}\sum_{i=1}^{n}\left\{ \left|\hat{\tau}\left(X_{i}\right)\right|l_{h}\left(u_{i}\right)\right\} +\lambda\sum_{j=1}^{p}\left(\left|\eta_{j}\right|/\left|\hat{\eta}_{int,j}^{\gamma}\right|\right),\label{eq:wsvmloss}
\end{eqnarray}
where $u_{i}=\left[2I\left\{ \hat{\tau}\left(X_{i}\right)>0\right\} -1\right]f\left(X_{i};\eta\right)$.
The subscript comes from the fact that $\hat{\eta}_{wsvm}$ can be
solved by running the WSVM algorithm with appropriate regularization
terms. A big advantage of using the hinge loss function is the low
computational cost because it is a convex programming problem. 

Alternatively, we can use the smoothed ramp loss function as the surrogate
function, and thus $\hat{\eta}_{dc}=\textrm{argmin}_{\eta}\hat{\mathcal{L}}_{\mathcal{F}}\left(\eta;\hat{\tau},\hat{\eta}_{int},l_{r}\right)$,
where
\begin{eqnarray}
\hat{\mathcal{L}}_{\mathcal{F}}\left(\eta;\hat{\tau},\hat{\eta}_{int},l_{r}\right) & = & \hat{\mathcal{R}}_{\mathcal{F}}\left(\eta;\hat{\tau},l_{r}\right)+J_{adplasso}\left(\eta;\hat{\eta}_{int}\right)\nonumber \\
 & = & \frac{1}{n}\sum_{i=1}^{n}\left\{ \left|\hat{\tau}\left(X_{i}\right)\right|l_{r}\left(u_{i}\right)\right\} +\lambda\sum_{j=1}^{p}\left(\left|\eta_{j}\right|/\left|\hat{\eta}_{int,j}^{\gamma}\right|\right).\label{eq:dcloss}
\end{eqnarray}
Unfortunately, $\hat{\mathcal{L}}_{\mathcal{F}}\left(\eta;\hat{\tau},\hat{\eta}_{int},l_{r}\right)$
is not a convex function and thus cannot be easily minimized by convex
programming algorithms. The solution is to use the d.c. algorithm
proposed by \citet{thi1997solving} that employs the fact that this non-convex loss function can be written as the difference
between two convex functions: $\hat{\mathcal{L}}_{\mathcal{F}}\left(\eta;\hat{\tau},\hat{\eta}_{int},l_{r}\right)=\mathcal{L}_{1}\left(\eta\right)-\mathcal{L}_{2}\left(\eta\right)$,
where
\begin{eqnarray*}
\mathcal{L}_{1}\left(\eta\right) & = & \frac{1}{n}\sum_{i=1}^{n}\left\{ \left|\hat{\tau}\left(X_{i}\right)\right|l_{1}\left(u_{i}\right)\right\} +\lambda\sum_{j=1}^{p}\left(\left|\eta_{j}\right|/\left|\hat{\eta}_{int,j}^{\gamma}\right|\right)\\
\mathcal{L}_{2}\left(\eta\right) & = & \frac{1}{n}\sum_{i=1}^{n}\left\{ \left|\hat{\tau}\left(X_{i}\right)\right|l_{0}\left(u_{i}\right)\right\} .
\end{eqnarray*}
The key idea of the d.c. algorithm for minimizing $\mathcal{L}\left(\eta\right)=\mathcal{L}_{1}\left(\eta\right)-\mathcal{L}_{2}\left(\eta\right)$
is to solve a convex subproblem iteratively. Denote $\nabla\mathcal{L}_{2}\left(\eta^{\left(t\right)}\right)=\left(\partial\mathcal{L}_{2}/\partial\eta_{1},\ldots,\partial\mathcal{L}_{2}/\partial\eta_{p}\right)\mid_{\eta=\eta^{(t)}}$
as the first order derivative evaluated at the $t$-th step estimate
$\eta^{(t)}$. The subproblem is then constructed as $\mathcal{L}_{1}\left(\eta\right)-\nabla\mathcal{L}_{2}\left(\eta^{\left(t\right)}\right)\eta$.
It is convex because adding a linear function does not affect convexity.
\citet{thi1997solving} proved that the minimizer of the convex subproblem
$\eta^{(t)}$ will converge to the minimizer of $\mathcal{L}\left(\eta\right)$.

\begin{algorithm}
\caption{The d.c. algorithm to minimize $\mathcal{L}\left(\eta\right)=\mathcal{L}_{1}\left(\eta\right)-\mathcal{L}_{2}\left(\eta\right)$}

\begin{algor}
\item [{{*}}] Set $\epsilon$ to be a small positive number as the tolerance
of error, say $\epsilon=10^{-5}$;
\item [{Initialize}] $\eta^{\left(0\right)}$;
\item [{while}] $\parallel\eta^{\left(t\right)}-\eta^{\left(t-1\right)}\parallel\leq\epsilon$
\begin{algor}
\item [{{*}}] $\eta^{(t+1)}=\textrm{argmin}_{\eta}\mathcal{L}_{1}\left(\eta\right)-\nabla\mathcal{L}_{2}\left(\eta^{\left(t\right)}\right)\eta$;
\end{algor}
\item [{endwhile}]~
\end{algor}
\end{algorithm}

In the ITR problem with smooth ramp loss function, because
\begin{eqnarray*}
\nabla\mathcal{L}_{2}\left(\eta^{\left(t\right)}\right)\eta & = & \frac{1}{n}\sum_{i=1}^{n}\left\{ \left|\hat{\tau}\left(X_{i}\right)\right|\frac{\partial l_{0}\left(u_{i}\right)}{\partial u_{i}}\frac{\partial u_{i}}{\partial\eta}\mid_{\eta=\eta^{\left(t\right)}}\right\} \eta\\
 & = & \frac{1}{n}\sum_{i=1}^{n}\left\{ \left|\hat{\tau}\left(X_{i}\right)\right|\frac{\partial l_{0}\left(u_{i}\right)}{\partial u_{i}}\mid_{u_{i}=u_{i}^{\left(t\right)}}\left[2I\left\{ \hat{\tau}\left(X_{i}\right)>0\right\} -1\right]X^{\T}\eta\right\} \\
 & = & \frac{1}{n}\sum_{i=1}^{n}\xi_{i}^{\left(t\right)}u_{i},
\end{eqnarray*}
where $\xi_{i}^{\left(t\right)}=\left|\hat{\tau}\left(X_{i}\right)\right|\partial l_{0}\left(u_{i}^{\left(t\right)}\right)/\partial u_{i}$
and $u_{i}^{\left(t\right)}=\left[2I\left\{ \hat{\tau}\left(X_{i}\right)>0\right\} -1\right]f\left(X_{i};\eta^{\left(t\right)}\right)$.
Therefore, the convex subproblem is
\begin{equation}
\eta^{\left(t+1\right)}=\underset{\eta}{\textrm{argmin}}\frac{1}{n}\sum_{i=1}^{n}\left\{ \left|\hat{\tau}\left(X_{i}\right)\right|l_{1}\left(u_{i}\right)-\xi_{i}^{\left(t\right)}u_{i}\right\} +\lambda\stackrel[j=1]{p}{\sum}\left(\left|\eta_{j}\right|/\left|\hat{\eta}_{int,j}^{\gamma}\right|\right),\label{eq:convexsub}
\end{equation}
which is not differentiable only at $0$. Various optimization algorithms prove effective in addressing this non-standard problem featuring L1-type regularization \citep{schmidt2007fast}. Interestingly, derivative-based methods like L-BFGS \citep{nocedal1980updating} have shown promise, even in scenarios where the problem lacks differentiability in certain areas \citep{guo2018nonsmooth}. However, the original convergence criterion $\parallel\eta^{\left(t\right)}-\eta^{\left(t-1\right)}\parallel\leq\epsilon$ might be overly strict, particularly in cases with high-dimensional $\eta$. Consequently, we opt to use the loss function directly as our criterion for halting the iteration. The loss function always provides a one-dimensional quantity, simplifying the convergence criterion.

\begin{algorithm}
\caption{The d.c. algorithm to minimize the loss function with smoothed ramp
loss and adaptive LASSO penalty}

\begin{algor}
\item [{{*}}] Set $\epsilon$ to be a small positive number as the tolerance
of error, say $\epsilon=10^{-5}$;
\item [{Initialize}] $\eta^{\left(0\right)}=\hat{\eta}_{int}$;
\item [{while}] $\left|\hat{\mathcal{L}}_{\mathcal{F}}\left(\eta^{\left(t\right)};\hat{\tau},\hat{\eta}_{int},l_{r}\right)-\hat{\mathcal{L}}_{\mathcal{F}}\left(\eta^{\left(t-1\right)};\hat{\tau},\hat{\eta}_{int},l_{r}\right)\right|\leq\epsilon$
\begin{algor}
\item [{{*}}] Update $u_{i}^{\left(t\right)}=\left[2I\left\{ \hat{\tau}\left(X_{i}\right)>0\right\} -1\right]f\left(X_{i};\eta^{\left(t\right)}\right)$;
\item [{{*}}] Update $\xi_{i}^{\left(t\right)}=\left|\hat{\tau}\left(X_{i}\right)\right|\partial l_{0}\left(u_{i}^{\left(t\right)}\right)/\partial u_{i}$;
\item [{{*}}] Update $\eta^{(t+1)}$ by solving the convex subproblem from
\eqref{eq:convexsub};
\end{algor}
\item [{endwhile}]~
\end{algor}
\end{algorithm}

\subsection{Main algorithm using cross validation\label{subsec:Main-algorithm}}

An essential consideration in applying our approach is the selection of the penalty parameter $\lambda$, that controls the degree of variable selection aggressiveness. When $\lambda$ is exceedingly large, the resulting ITR becomes trivial, assigning all subjects to the same treatment. Conversely, a very small $\lambda$ yields an ITR close to the optimal but sacrifices interpretability. Our proposed approach involves leveraging cross-validation (CV) to tune $\lambda$. We partition the data into $K$ folds\textemdash commonly, $K=5$ or $10$\textemdash using in turn one fold as the test set and the remaining $K-1$ folds as the training set.

For each candidate $\lambda$, we minimize the loss function (\eqref{eq:wsvmloss} or \eqref{eq:dcloss}) using WSVM or the d.c. algorithm, respectively, to estimate the policy coefficient in the training set. Subsequently, $\hat{\eta}$ is evaluated using the value function estimated from \eqref{eq:valueAIPW} on the test set. The overall performance for each $\lambda$ is obtained by averaging across all $K$ folds, ensuring each fold serves as the test set.

We present two methods for selecting an appropriate $\lambda$: $\lambda_{\text{min}}$, having the highest estimated value on average, and $\lambda_{1\text{se}}$, the largest $\lambda$ among those with estimated values within one standard error of $\lambda_{\text{min}}$. Notably, $\lambda_{1\text{se}}$, being no less than $\lambda_{\text{min}}$, tends to be more aggressive in eliminating unimportant variables in the IITR.

Once $\lambda$ is chosen, the complete minimization process is conducted afresh using the full dataset to maximize accuracy. We identify unimportant variables based on $\hat{\eta}_{\textrm{full}}$\textemdash for instance, eliminating variables with an absolute magnitude less than $0.1 \times$ the maximum absolute coefficient. Subsequently, the algorithm is refit using the selected variables and $\lambda=0$. The main algorithm progresses as follows:

\begin{algorithm}

\caption{IITR Algorithm with Adaptive LASSO Using Cross-Validation}
\label{alg:main}
\begin{algor}
\item [{{*}}] Normalize covariates such that each covariate has mean zero
and standard deviation one;
\item [{{*}}] Split the dataset into $K$ folds;
\item [{for}] $i$ in $1:K$
\begin{algor}
\item [{{*}}] Use the $i$-th fold as the test set and other folds as the
training set;
\item [{{*}}] Using the training set, estimate the contrast function from
\eqref{eq:iteAIPW} via AIPW or AIPWSL; 
\item [{{*}}] Run WSVM or d.c. algorithm to obtain an initial estimate
$\hat{\eta}_{int}$ by minimizing \eqref{eq:wsvmloss} or \eqref{eq:dcloss}
with $\lambda=0$; 
\item [{for}] $\lambda$ in a sequence of pre-specified penalty parameters
$\lambda_{1},...,\lambda_{L}$; 
\begin{algor}
\item [{{*}}] Estimate the coefficient $\hat{\eta}$ by minimizing \eqref{eq:wsvmloss}
or \eqref{eq:dcloss} using WSVM or d.c. algorithm with penalty parameter
$\lambda$ and initial estimate $\hat{\eta}_{int}$; 
\item [{{*}}] Using the test set, evaluate the performance by estimating
the value function of $\hat{\eta}$ from \eqref{eq:valueAIPW} via
AIPW or AIPWSL;
\end{algor}
\item [{endfor}]~
\end{algor}
\item [{endfor}]~
\item [{{*}}] Evaluate overall performance for $\lambda_{1},...,\lambda_{L}$
by averaging the estimated values over $K$ folds;
\item [{{*}}] Using the full dataset, estimate the contrast function from
\eqref{eq:iteAIPW} via AIPW or AIPWSL;
\item [{{*}}] Using either $\lambda_{min}$ or $\lambda_{1se}$, estimate
the coefficients $\hat{\eta}_{\textrm{full}}$ using WSVM or d.c.
algorithm by minimizing \eqref{eq:wsvmloss} or \eqref{eq:dcloss}; 
\item [{{*}}] Remove unimportant variables based on $\hat{\eta}_{\textrm{full}}$,
e.g., remove variables with absolute magnitude less than $0.1\times$max
absolute coefficient;
\item [{{*}}] Refit the algorithm by minimizing \eqref{eq:wsvmloss} or
\eqref{eq:dcloss} using WSVM or d.c. algorithm with selected variables
and $\lambda=0$. 
\end{algor}
\end{algorithm}

\subsection{Complementary analysis procedure of flexible variable selection\label{subsec:Complementary-analysis}}

Sometimes, researchers would like to obtain a simple policy with pre-specified
limited number of variables, or they would like to get a sense of
the number of variables to be kept in the policy. Also, some variables
in the observed dataset may be expensive and difficult to collect
in practice and thus are better not to be included in the policy.
Existing methods and the main algorithm in Section \eqref{subsec:Main-algorithm}
may offer an optimal interpretable ITR constructed by some selected
variables, but the number of variables are implicitly determined by
the magnitude of penalty parameter and cannot be easily tuned by researchers.
To deal with this problem, we propose a complementary analysis procedure
to select variables flexibly. The key idea is to rank the importance
of $p_{selected}$ variables based on the absolute magnitude of $\hat{\eta}_{\textrm{full}}$
from Algorithm \ref{alg:main}, where $p_{selected}$ is the number
of variables that are feasible to be included in the policy. Because
variables have been normalized, larger coefficient in the policy vector
implies more importancy. On the other hand, variables with coefficients
closed to zero are unimportant and should be removed from the property
of adaptive LASSO algorithm. Inspired by this, we use the $k$ most
important variables to construct the optimal policy by minimizing
\eqref{eq:wsvmloss} or \eqref{eq:dcloss} with $\lambda=0$ and the
$k$ selected variables, where $k$ goes from $1$ to $p_{selected}$.
We then evaluate the performance of each policy by estimating the
corresponding value function from \eqref{eq:valueAIPW} via AIPW or
AIPWSL, and the values are plotted in a graph, where $x$ axis is
the number of variables in the policy ranged from $1$ to $p$ and
$y$ axis is the corresponding value. We further extend the number
of variables in the policy to $0$, implying the trivial policies
that assign everyone to the treatment group or everyone to the control
group. We evaluate the values of these two policies and choose the
larger one as the value of the $0$-variable policy. The confidence
intervals of the policies are also calculated and form a confidence
band for the value function. This complementary analysis procedure
is summarized in Algorithm \ref{alg:complem}.

\begin{algorithm}
\caption{Complementary analysis procedure of flexible variable selection}
\label{alg:complem}
\begin{algor}
\item [{{*}}] Using the full dataset, estimate the contrast function from
\eqref{eq:iteAIPW} via AIPW or AIPWSL;
\item [{{*}}] Ignore the variables that should not be included in the policy,
and denote the number of variables remained as $p_{selected}$;
\item [{{*}}] Rank the importance of variables based on the absolute magnitude
of $\hat{\eta}_{\textrm{full}}$ from Algorithm \ref{alg:main};
\item [{{*}}] Evaluate the value function of the trivial policies, i.e.,
$\hat{\eta}=\left(\pm1,0,\ldots,0\right)$, from \eqref{eq:valueAIPW}
via AIPW or AIPWSL, and record the larger one as the value for $k=0$;
\item [{for}] $k$ in $1:p_{selected}$
\begin{algor}
\item [{{*}}] Estimate the optimal $k$-variable policy $\hat{\eta}_{k}$
by minimizing \eqref{eq:wsvmloss} or \eqref{eq:dcloss} using WSVM
or d.c. algorithm with the $k$ most important variables and $\lambda=0$;
\item [{{*}}] Evaluate the value function of $\hat{\eta}_{k}$ from \eqref{eq:valueAIPW}
via AIPW or AIPWSL, and calculate its confidence interval based on
\eqref{eq:valueAIPWvar};
\end{algor}
\item [{endfor}]~
\item [{{*}}] Plot the graph of value function, where $x$ axis is the
number of variables in the policy ranged from $0$ to $p_{selected}$,
and $y$ axis is the corresponding value;
\item [{{*}}] Plot the confidence band based on the obtained confidence
intervals.
\end{algor}
\end{algorithm}

We emphasize the importance of this graph of value function because
it can offer a few important insights to choose the appropriate number
of variables kept in the ITR. For example, the value plot will be
an increasing trend because more variables imply a more complex and
informative policy. However, including unimportant variables may not
significantly increase the value if the additional information is
redundant. Thus, we recommend to choose the number of variables where
the trend changes from steep to flat so that the information included
is most efficient. This selection process is subjective and flexible
to the researchers, and researchers can also relate this result to
their prior knowledge. Alternatively, researchers may also select
the number of variables if the corresponding value first exceeds a
scientifically reasonable thereshold or statistically significantly
better than the trivial policies. We offer a detailed explaination
in the following real data application in the following section to
illustrate this idea.

\section{Real data application\label{sec:Realdata}}

In this section, we applied our proposed methods to the TRIUMPH study (Preventive Treatment of Migraine: Outcomes for Patients in Real-World Healthcare Systems) \citep{lipton2025treatment}. The goal to show how the proposed ITR using adaptive LASSO to optimize the treatment regimen between galcanezumab and other preventive oral migraine treatments (TOMP). TRIUMPH is an ongoing, 24-month, prospective, multicenter, international, observational study of patients with migraine at the time of initiating or switching to pharmacologic treatment for migraine prevention (European Network of Centers for Pharmacoepidemiology and Pharmacovigilance identifier: EUPAS33068). The study enrolled patients from the US, Japan, Germany, Italy, Spain, United Kingdom, and United Arab Emirates. For illustration purpose, this analysis used data collected between February 25, 2020, and February 9, 2023. This analysis compares the 3-month treatment effectiveness of galcanezumab versus TOMP as the Individualized Treatment Effect. Adult patients with a diagnosis of migraine were enrolled at the time when they were prescribed a new pharmacologic migraine preventive treatment (index drug). For the current analysis, patients had to report $\geq4$ migraine headache days in the 30 days preceding study start and taken galcanezumab (at the approved dose/regimen) or TOMP as the index drug. Based on the treatment initiated, 2190 patients were grouped into galcanezumab (initiating galcanezumab, including the loading dose per label, 884 pateints) and standard of care (initiating select medications within the drug classes of anticonvulsants, tricyclic antidepressants, beta-blockers, calcium channel blockers, or angiotensin II receptor antagonists) cohorts in this 3-month assessment. Patients receiving other CGRP monoclonal antibodies (mAbs), botulinum toxin, or other locally approved medications are also considered in the group of standard treatments. This study assessed the treatment effectiveness of galcanezumab versus standard of care by measuring the change in monthly migraine headache days from patient responses recorded by physicians at the 3-month visit. The primary outcome was the proportion of patients with a clinically meaningful response at 3 months, combining all patients using different thresholds per migraine type, namely a reduction from baseline in monthly migraine headache days of $\geq50$\% for patients with episodic migraine and $\geq30$\% for patients with chronic migraine.16, 17 Non responder imputation (NRI) was applied to the response variable, meaning patients were considered non-responders if they discontinued the study, were lost to follow-up before the 3-month visit, or missed the 3 month visit window. Although patients were expected to remain in the study if they discontinued their index drug, the NRI method was implemented as a conservative response estimate to account for patients who discontinued the study for any reason, especially due to lack of efficacy or poor tolerability.

Before applying our method, we imputed the missing values in the dataset
using the R package ``mice''. Because multiple imputation cannot
be directly applied to the ITR setting, we made single imputation
by averaging the multiple imputed values. For ethical reasons, we
removed some variables from the policy, including ethnicity, type
of center, type of provider, reduced work Flag, type of stage 1 visit,
new preventive treatment added to previous, employed flag, number
of comorbidities at baseline. However, we could still use these variables
to estimate the ITE for the subjects to improve accuracy thanks to
the flexibility of the algorithm. 

\begin{figure}
\caption{Estimated absolute coefficients of the optimal ITRs derived from the
IITR algorithm. The coefficients were ranked and standardized by dividing them by
the largest absolute value. \label{fig:Estimated-absolute-coefficients}}

\centering{}\includegraphics[width=\textwidth]{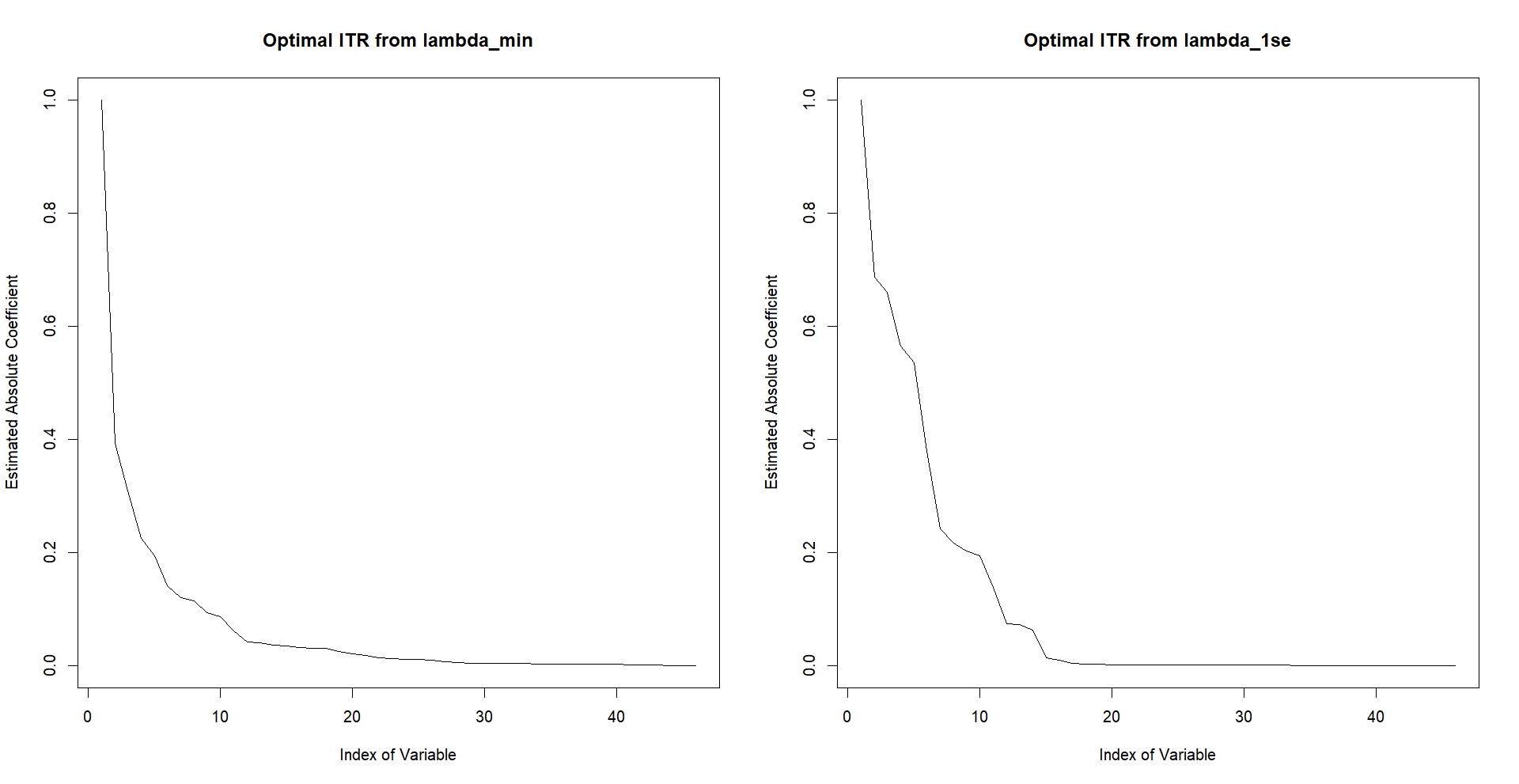}
\end{figure}

We applied our IITR algorithm to the TRIUMPH dataset using the 5-fold
cross-validation and d.c. algorithm. We also tried the version with
WSVM, but the performance was not better than the d.c. algorithm and
thus we don't show that result here. Nusiance functions were estimated
via SuperLearner with base functions including penalized regression,
random forest, generalized additive model, gradient boosting, and
neural network. The tuning parameter $\lambda$ was selected from
a geometrically equally spaced sequence from $10^{-3}$ to $10^{7}$
with length $500$, and the algorithm chose $\lambda_{min}=955$ and
$\lambda_{1se}=7\times10^{6}$. To simplify notations, we denote the
two ITRs as $\hat{\eta}_{min}$ and $\hat{\eta}_{1se}$. Figure \ref{fig:Estimated-absolute-coefficients}
showed the estimated absolute coefficients of the optimal ITRs derived
from the IITR algorithm. The coefficients were ranked and standardized
by dividing the largest absolute value. Because covariates had been
standardized before running the algorithm, the importance of the variables
could be compared based on their corresponding coeffcients. The coefficients
of $\hat{\eta}_{1se}$ decreased faster to zero compared to $\hat{\eta}_{min}$,
implying that larger penalty parameters removed more unimportant variables
from the policy. This can be seen from the number of variables with
standardized coefficients larger than $0.01$: $22$ variables are
remained in $\hat{\eta}_{min}$, while $14$ variables are kept in
$\hat{\eta}_{1se}$. As a result, overfitting may be prevented in
the more agressive policy: the estimated value of $\hat{\eta}_{min}$
is $0.98$, whereas the estimated value of $\hat{\eta}_{1se}$ is
$1.66$. To improve performance, we refit the algorithm by only including
the $10$ variables with standardized coefficients from $\hat{\eta}_{1se}$
larger than $0.1$ in the policy. The estimated value of the refitted
ITR is $2.72$.

\begin{figure}
\caption{Complementary analysis results of flexible variable selection based
on $\hat{\eta}_{1se}$ for the TRIUMPH study. Dotted lines correspond
to the $95\%$ confidence bound. ITR with zero variable corresponds
to the best trivial policy that assigns everyone to the same treatment
group. In the TRIUMPH study, it is giving galcanezumab to all the
patients, and its estimated value is 0.47. \label{fig:Complementary-analysis-results}}

\centering{}\includegraphics[width=\textwidth]{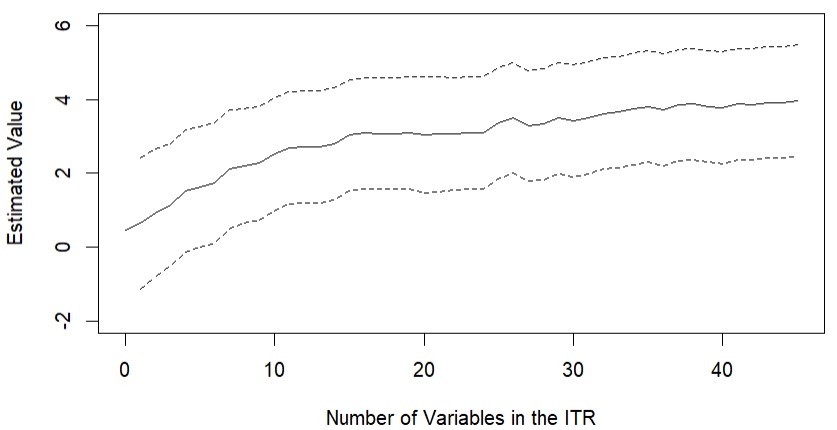}
\end{figure}

The number of variables kept in the reduced policy depends on the
selection criterion, while it is unclear how to choose a reasonable
value. To have a better understanding of how much information we could
obtain by including more variables into the ITR, we proceed with the
complementary analysis procedure of flexible variable selection based
on $\hat{\eta}_{1se}$. Figure \ref{fig:Complementary-analysis-results}
shows the results from the complementary analysis procedure by gradually
adding variables into the ITR based on their correpsonding absolute
coefficients in $\hat{\eta}_{1se}$. The estimated value of the ITR
increases when the number of variables in the ITR increases, and the
trend is rapid when the number of variables is small. However, the
increasing trend flattens out after including a sufficient number of important variables.
We can target on the place where the trend becomes flat. From the
plot, it happens when the number of variables equals to $11$ and
$15$. Thus, it is reasonable and efficient to choose an ITR with
$11$ or $15$ variables. Table \ref{tab:List-of-important} shows
the list of names of the most important $15$ variables and the estimated
values of the corresponding policies with confidence intervals. Physicians
may apply the more comprehensive ITR if additional information such
as PGIS score is obtainable. Moreover, ITR with zero variable in Figure
\ref{fig:Complementary-analysis-results} corresponds to the best
trivial policy that assigns everyone to the same treatment group.
In the TRIUMPH study, it is giving galcanezumab to all the patients,
and the estimated value of this policy is 0.47. If physicians would
like to use a more interpretable ITR with less variables, a $7$-variable
policy is also a reasonable choice because it performs significantly
better than the trivial policy.

\begin{table}
\caption{List of important variable names in the complementary analysis based
on $\hat{\eta}_{1se}$ for the TRIUMPH study, ranking from the most
important to the least important. The values of the ITRs and the corresponding
$95\%$ confidence intervals are estimated by AIPW with SuperLearner.
\label{tab:List-of-important}}

\begin{centering}
\resizebox{\columnwidth}{!}{%
\begin{tabular}{|c|c|c|c|}
\hline 
Index & i-th important variable & Estimated value for i-th regime & Confidence inverval for the value\tabularnewline
\hline 
\hline 
1 & Acid reflus / gerd flag & 0.65 & (-1.12,2.42)\tabularnewline
\hline 
2 & AGE & 0.93 & (-0.81,2.66)\tabularnewline
\hline 
3 & Nausea flag & 1.14 & (-0.51,2.79)\tabularnewline
\hline 
4 & Vomiting flag & 1.53 & (-0.12,3.17)\tabularnewline
\hline 
5 & Race\_black & 1.63 & (-0.01,3.25)\tabularnewline
\hline 
6 & Sex\_male & 1.74 & (0.12,3.36)\tabularnewline
\hline 
7 & Midas score & 2.11 & (0.50,3.73)\tabularnewline
\hline 
8 & Family history of migraine & 2.20 & (0.65,3.75)\tabularnewline
\hline 
9 & OPIOID/BARB baseline flag & 2.27 & (0.74,3.81)\tabularnewline
\hline 
10 & Number of days migraine at baseline & 2.52 & (0.99,4.05)\tabularnewline
\hline 
11 & Number of prior acute treatments failed & 2.70 & (1.19,4.22)\tabularnewline
\hline 
12 & Rebound headache flag & 2.72 & (1.21,4.23)\tabularnewline
\hline 
13 & PGIS score & 2.73 & (1.21,4.24)\tabularnewline
\hline 
14 & Photophobia flag & 2.80 & (1.28,4.31)\tabularnewline
\hline 
15 & Asthma flag & 3.03 & (1.52,4.54)\tabularnewline
\hline 
\end{tabular}
}
\par\end{centering}
\end{table}

\section{Simulations\label{sec:simulation}}

In this section, we conducted a simulation study to compare the performance
of our IITR algorithm to the existing state-of-the-art methods, including
causal forest \citep{hahn2020bayesian}, and R-learning \citep{nie2021quasi}.
In practice, the underlying models of treatment assignment and potential
outcome may be complicated, while the mechanism of treatment effect
may be simpler. To reproduce this phenomenon, we designed a simulation
setting with $20$ covariates following independent standard normal
distributions, while the true model of ITR is a second-order polynomial
of only two variables. Our goal is to estimate the optimal linear
policy which should only depends on those two variables. We used generalized
linear models to generate the propensity score and potential outcomes
for the control group. We generated $3000$ subjects in the training set
to estimate the ITR and $1000$ subjects under the same distribution
to evaluate the performance. The experiments were replicated for $1000$
times.

We used the 5-fold cross-validation and d.c. algorithm to operate
our IITR algorithm. The performance of the algorithm with WSVM was
very similar and thus we omitted those results. We used AIPW to estimate
the ITE, and first-order generalized linear models were used to estimate
the nuisance functions in the algorithm. The tuning parameter $\lambda$
was selected from a geometrically equally spaced sequence from $10^{-4}$
to $10$ with length $20$. Variables with absolute coefficient less
than $0.01\times$max absolute coefficient were removed from the policy,
and an interpretable ITR was estimated using the remained covariates.
Causal forest was implemented using the R package
``grf'' and R-learning was implemented using the R package ``rlearner''.

\begin{figure}
\caption{Simulation results of the comparison among IITR, causal forest, and
R-learning. The left and right panel show the values and correct classification
rates of the estimated ITRs, respectively. \label{fig:sim-results}}

\centering{}\includegraphics[width=\textwidth]{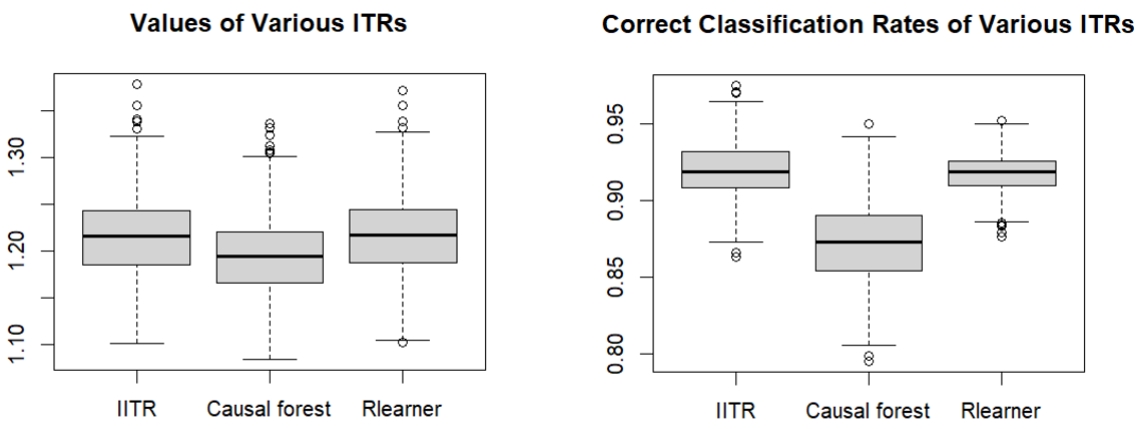}
\end{figure}

Figure \ref{fig:sim-results} shows the simulation results of the
comparison among the three algorithms, including the values and correct
classification rates of the estimated ITRs. It can be seen that the
performance of IITR is superior to that of causal forest and comparable to
R-learning. For the correct classification rate, the variation of
IITR is larger than R-learning. The higher rate may be explained by the
fact that variable selection reduces noise and makes the estimated
policy closer to the truth. On the other hand, the lower rate may occur 
because some important variables could be accidentally removed. Nevertheless,
the performance of the IITR algorithm can be improved by carefully
tuning and operating the complementary analysis procedure of flexible
variable selection, while R-learning does not have this flexibility
and interpretability. 

\section{Discussion\label{sec:Discussion}}

In this work, we propose a novel framework for estimating optimal and interpretable ITR in high-dimensional settings. By formulating ITR estimation as a classification problem and leveraging adaptive LASSO for variable selection, our method achieves a balance between predictive accuracy and model interpretability. A key advantage of our approach is the flexibility to adjust this balance by modifying the variables included in the policy, guided by the visualization of policy coefficients and the value function. This complementary analysis can be easily integrated with domain knowledge to scientifically select relevant variables or efficiency thresholds.

There are several areas where future work could build on this approach. First, while linear ITRs are widely used for their interpretability, some practitioners may prefer alternatives like tree-based models that more closely mirror human decision-making. However, optimizing the value or risk function under tree-based ITRs remains a challenging problem. Second, our current method assumes fully observed outcomes, but in practice, missing data and censored outcomes can be common due to early dropout. Extending the algorithm to handle censored or missing outcomes in survival analysis contexts would increase its applicability. Finally, further investigation into the theoretical properties of the method, including consistency and convergence rates in high-dimensional settings, would provide stronger guarantees and deeper insight into its performance.

\bibliographystyle{Chicago}
\bibliography{ci}

\end{document}